\documentclass[fleqn,usenatbib]{mnras}

\usepackage{newtxtext,newtxmath}
\usepackage{graphics}
\usepackage{epsfig}	
\usepackage{amsmath}	

\usepackage{amssymb}	
\usepackage[T1]{fontenc}

\DeclareRobustCommand{\VAN}[3]{#2}
\let\VANthebibliography\thebibliography
\def\thebibliography{\DeclareRobustCommand{\VAN}[3]{##3}\VANthebibliography}

\newcommand{\dmmw}{{\rm DM}_{\rm MW}}

\title[81 new candidate FRBs]{81 New Candidate Fast Radio Bursts in Parkes Archive}

\author[X. Yang, S.-B. Zhang et al.]{
X. Yang$^{1,2}$, S.-B. Zhang$^{1}\thanks{E-mail: sbzhang@pmo.ac.cn}$, J.-S. Wang$^{3}$, G. Hobbs$^{4}$, T.-R. Sun$^{1,2}$, R. N. Manchester$^{4}$, 
\newauthor J.-J. Geng$^{1}$, C. J. Russell$^{5}$, R. Luo$^{4}$, Z.-F.Tang$^{1,2}$, C. Wang$^{6}$,  J.-J. Wei$^{1}$, L. Staveley-Smith$^{7,8}$, 
\newauthor S. Dai$^{9}$, Y. Li$^{1}$, Y.-Y. Yang$^{10}$, X.-F. Wu$^{1}\thanks{E-mail: xfwu@pmo.ac.cn}$
\\
$^{1}$ Purple Mountain Observatory, Chinese Academy of Sciences, Nanjing 210023, China\\
$^{2}$ School of Astronomy and Space Sciences, University of Science and Technology of China, Hefei 230026, China\\
$^{3}$ Tsung-Dao Lee Institute, Shanghai Jiao Tong University, Shanghai 200240, China\\
$^{4}$ CSIRO Space and Astronomy, Australia Telescope National Facility, PO Box 76, Epping, NSW 1710, Australia\\
$^{5}$ CSIRO Scientific Computing Services, Australian Technology Park, Locked Bag 9013, Alexandria, NSW 1435, Australia\\
$^{6}$ CSIRO Data61, Australian Technology Park, Locked Bag 9013, Alexandria, NSW 1435, Australia\\
$^{7}$ International Centre for Radio Astronomy Research, University of Western Australia, Crawley, WA 6009, Australia\\
$^{8}$ ARC Centre of Excellence for All Sky Astrophysics in 3 Dimensions (ASTRO 3D)\\
$^{9}$ Western Sydney University, Locked Bag 1797, Penrith South DC, NSW 1797, Australia\\
$^{10}$ School of Physics and Electronic Science, Guizhou Education University, Guiyang 550018, China\\
}

\date{Accepted 2021 July 29. Received 2021 July 29; in original form 2020 May 3}

\pubyear{2021}

\begin{document}
\label{firstpage}
\pagerange{\pageref{firstpage}--\pageref{lastpage}}
\maketitle

\begin{abstract}
We have searched for weak fast radio burst (FRB) events using a database containing 568,736,756 transient events detected using the Parkes radio telescope between 1997 and 2001.  In order to classify these pulses, and to identify likely FRB candidates, we used a machine learning algorithm based on ResNet.  We identified 81 new candidate FRBs and provide details of their positions, event times, and dispersion measures. These events were detected in only one beam of the Parkes multibeam receiver.  We used a relatively low S/N cutoff threshold when selecting these bursts and some have dispersion measures only slightly exceeding the expected Galactic contribution.  We therefore present these candidate FRBs as a guide for follow-up observations in the search for repeating FRBs.  
\end{abstract}

\begin{keywords}
fast radio bursts < Transients, methods: data analysis < Astronomical instrumentation, methods, and techniques, Astronomical Data bases
\end{keywords}


\section{Introduction}
Fast Radio Bursts (FRBs) are bright radio pulses of short duration which range from microseconds to milliseconds. Since \citet{Lorimer:2007qn} reported the first FRB in an archival observation from the Parkes 64-m-diameter radio telescope, hundreds of FRBs have been published\footnote{The FRB online Catalogue is available from \url{www.wis-tns.org}}
. The majority of these events are seen as one-off bursts, however 20 are so far identified as producing repeated pulses and are hence referred to as repeating FRBs~\citep{Spitler:2016dmz,CHIME19_2r,CHIME19_8r,Kumar19,Luo20,Geng_2021}. 
FRBs are generally defined as pulse sources with dispersion measures (DMs) greater than can be attributed to our Galaxy.
A handful of FRBs have now been localised to their host galaxies \citep[e.g.][]{Chatterjee17,Bannister19,DSA10,Prochaska19,askap_frb190711,Marcote20}. 
Recently, an extremely intense radio burst (FRB~200428) was detected from a Galactic magnetar, SGR~1935+2154, with an energy only $\sim30$ times weaker than the weakest extragalactic FRBs
\citep{frb200428,STARE2}. 
This discovery confirmed that FRBs can be produced by magnetars.
Combined with the jointly detected X-ray burst \citep{HXMT2020,INTEGRAL2020,Konus2020}, these results shed light on the FRB radiation mechanism (see \citealt{Wang2020b} for a review on this event, and \citealt{2020zhangbing,Xiao2021} for general reviews).

Most of the current FRB search software packages, such as \emph{\sc SIGPROC}\footnote{\url{http://sigproc.sourceforge.net/}},  \emph{\sc PRESTO}\footnote{\url{https://www.cv.nrao.edu/~sransom/presto/}} \citep[]{presto} and \emph{\sc HEIMDALL}\footnote{\url{https://sourceforge.net/projects/heimdall-astro/}}, are based on the single pulse search algorithm that was originally developed to find pulsars. 
In these pipelines, FRBs are found by searching a wide range of pulse widths and DM trials. 
To claim a new FRB detection, several criteria are applied, including thresholds on the minimum DM, the burst signal to noise ratio (S/N), and the maximum number of beams in which the source was detected~\citep[e.g.][]{Champion16,Bhandari18}. 

The DM threshold is based on the modelled Galactic DM contribution along the line of sight using the NE2001~\citep{Cordes02} or YMW16~\citep{Yao2017} electron-density models. 
The S/N cut-off is usually set between $6$ and $10$ in order to limit the number of false-positive candidates ~\citep{Foster18}. 
The majority of the published FRBs have a S/N above or close to 10.  This threshold value depends on the specific data acquisitions and processing pipelines~\citep{Zhang_2020,Parent20}. 
In this work we also adopt this threshold, namely FRBs are claimed as being detected if they are observed with a S/N $\geq10$, otherwise we refer to them as ``candidate FRBs''. 
Recently, two candidate FRBs with S/N of 8.4 and 8.1 were claimed to be real and extragalactic after comparison with other confirmed astrophysical candidates generated by the process of~\citet{Patel18} and~\citet{Parent20}. \citet{SNlow6} discussed a candidate FRB found along the M82 direction with S/N $\simeq6$.
For repeating FRBs, smaller S/N thresholds are  usually applied ~\citep{Spitler16,Gajjar18,CHIME19_8r}, and candidates with S/N as low as 6.4 have been presented~\citep{Gajjar18}. 


Assuming normally distributed noise, a false-positive detection with relatively high S/N would be highly improbable \citep{Zhang_2020,Parent20}. This value depends on the specific data acquisition and processing pipelines. For our work, $\sim 0.3$ false-positive detections with S/N  $\ge$ 7.5 are expected to be generated in the entire data sets that we have searched. Detailed calculations are presented in Section~\ref{sec:Dis}.
%
 The threshold based on the maximum number of detected beams in which the source is detected in is primarily to reduce the false-positive detections caused by radio frequency interference (RFI), which often is seen in all or many beams of a multibeam receiver.

%

In principle, as RFI is complex, variable, and hard to quantify, all the candidate pulses obtained by the search pipelines should be viewed as a dynamic spectrum by eye. However, with larger and larger data volumes this is becoming impossible. 
In \citet{Zhang_2020}, we reprocessed all the observations that were carried out during the first four years of the Parkes Multibeam Receiver. A total of 568,736,756 pulses with a S/N larger than seven were obtained and recorded in a single pulse database, known as the Parkes Transient Database (PTD). 
In our previous work we applied a selection cut for events with S/N $\geq8$ in the PTD~\citep[see usage details in][]{Zhang_2020} and identified four published FRBs (FRBs~010125, 010312, 010621 and 010724), as well as a new FRB (FRB~010305) with S/N values of 17.9, 11.0, 15.8, 32.0 and 10.2, respectively. 
In this current work, we mine the database further in order to identify the weaker candidate FRBs.

Machine Learning (ML) algorithms are playing  an important roles in various research areas, including in the search for FRBs. 
\citet{wagstaff} and \citet{foster_ml} both introduced a random forest classifier to their FRB searching pipelines, while \citet{connor} applied a wide deep learning net to extract features from four different figures and catch true single pulses from large numbers of candidates. 
\citet{stz1748} detected five new FRBs using a real-time FRB detection system which included the use of a random forest algorithm. Recently, \cite{FETCH} developed a Python package (\texttt{FETCH}) comprised of 11 deep-learning models, and this package can detect all FRBs with S/N values above 10 in the ASKAP and Parkes data.
%
Motivated by such research, we applied an ML algorithm to the PTD. We chose a deep residual network (ResNet) model, which is commonly used in computer vision tasks ~\citep{heresnet}.
%
%

Candidate FRBs provide pointing directions to enable sensitive, but narrow-beam, telescopes such as the Five-hundred-meter Aperture Spherical radio Telescope (FAST) to search for more repeating events. However, we note that the positions of these candidates are only poorly determined to date. Of course, the detection of any new FRBs is of great astrophysical importance as relatively few are currently known and studied. 

%

In Section \ref{sec:data}, we describe our new pipeline, and apply it to the PTD. We describe the properties of the detected 81 candidate FRBs in Section~\ref{sec:result}. We discuss our results in Section~\ref{sec:Dis}. 

\section{Implementation}
\label{sec:data}

\begin{figure*}
    \begin{center}
    \begin{tabular}{ccc}
    \includegraphics[width=5.5cm]{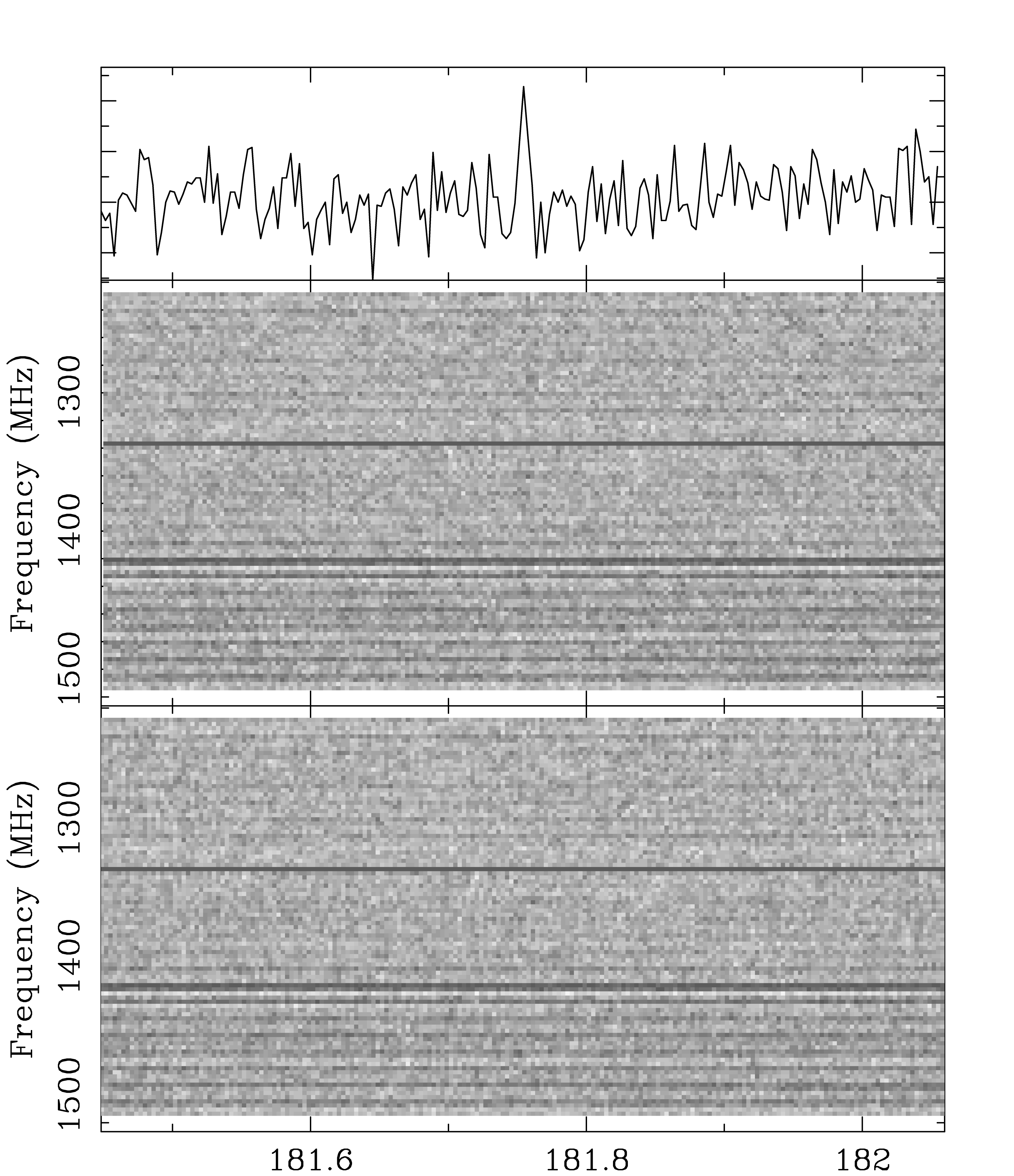} & 
    \includegraphics[width=5.5cm]{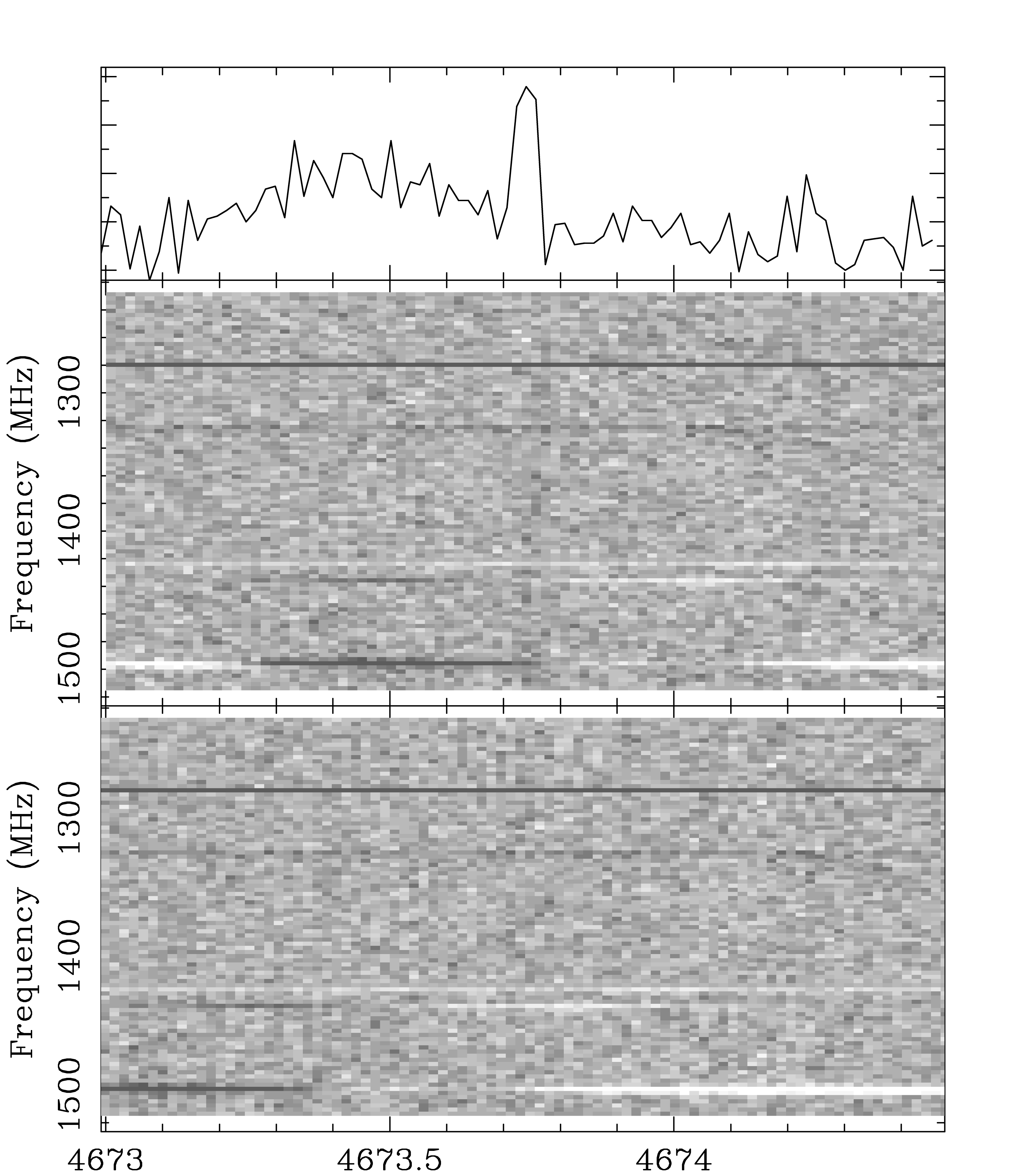} & 
    \includegraphics[width=5.5cm]{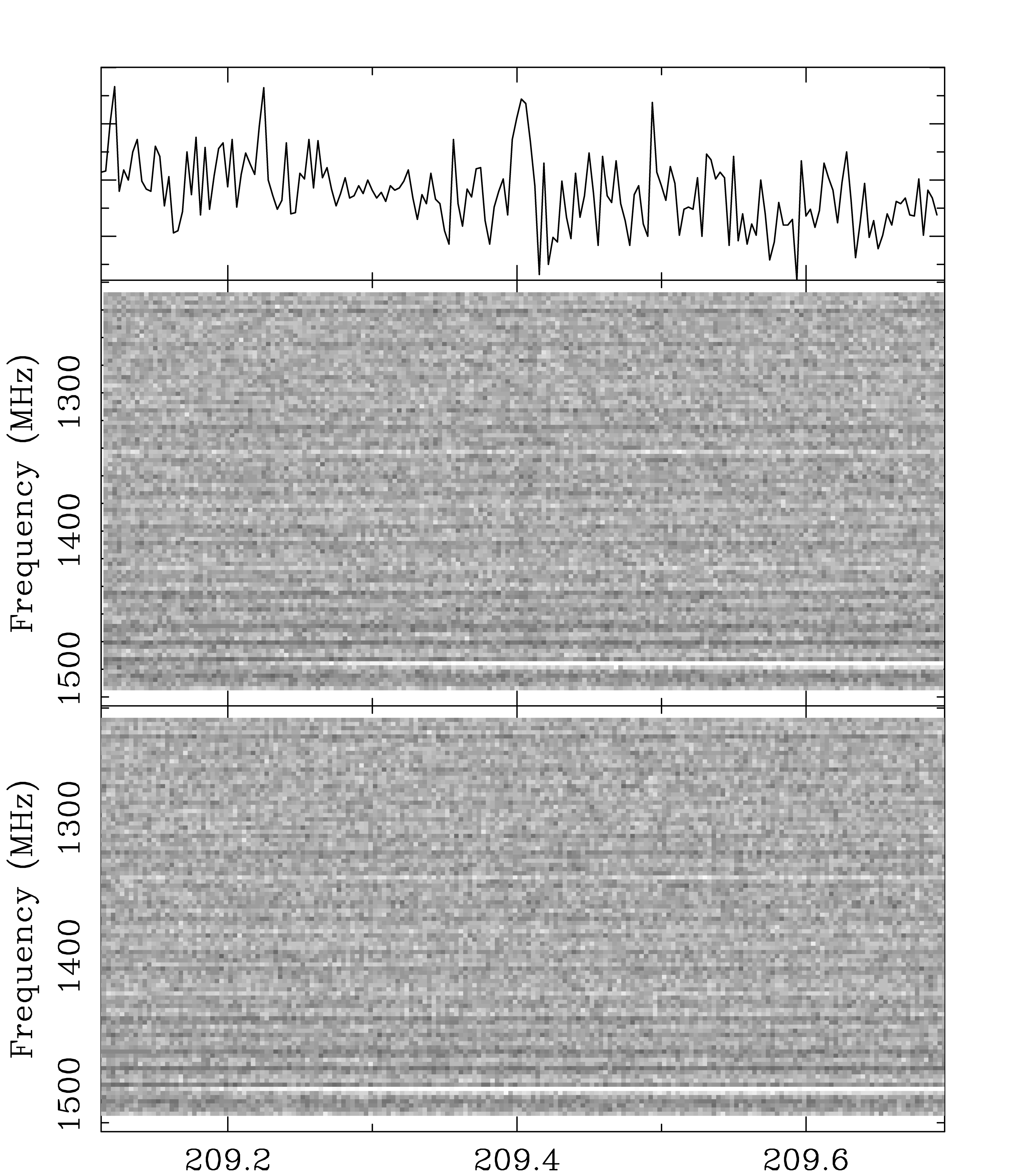}\\
    \includegraphics[width=5.5cm]{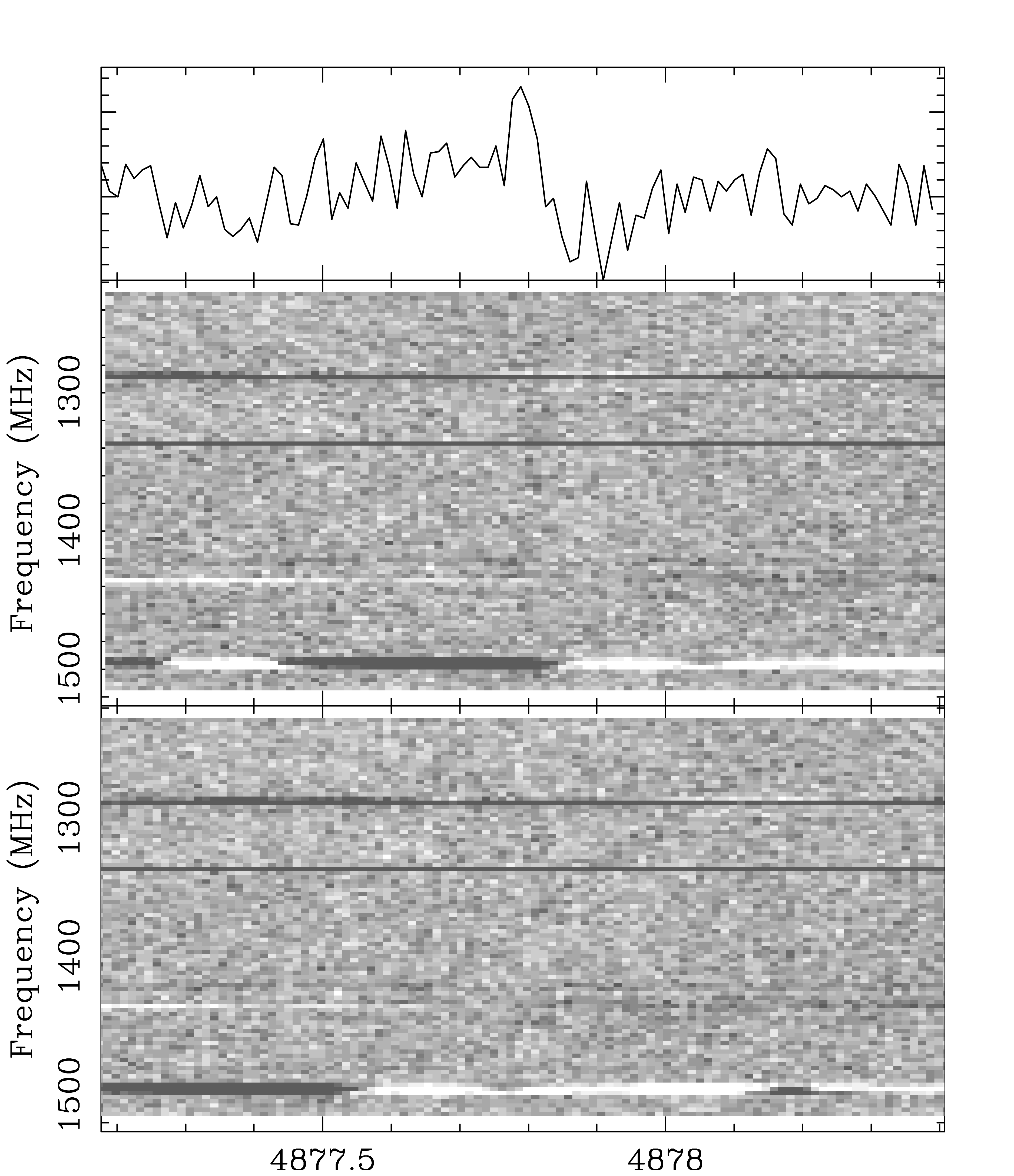} & 
    \includegraphics[width=5.5cm]{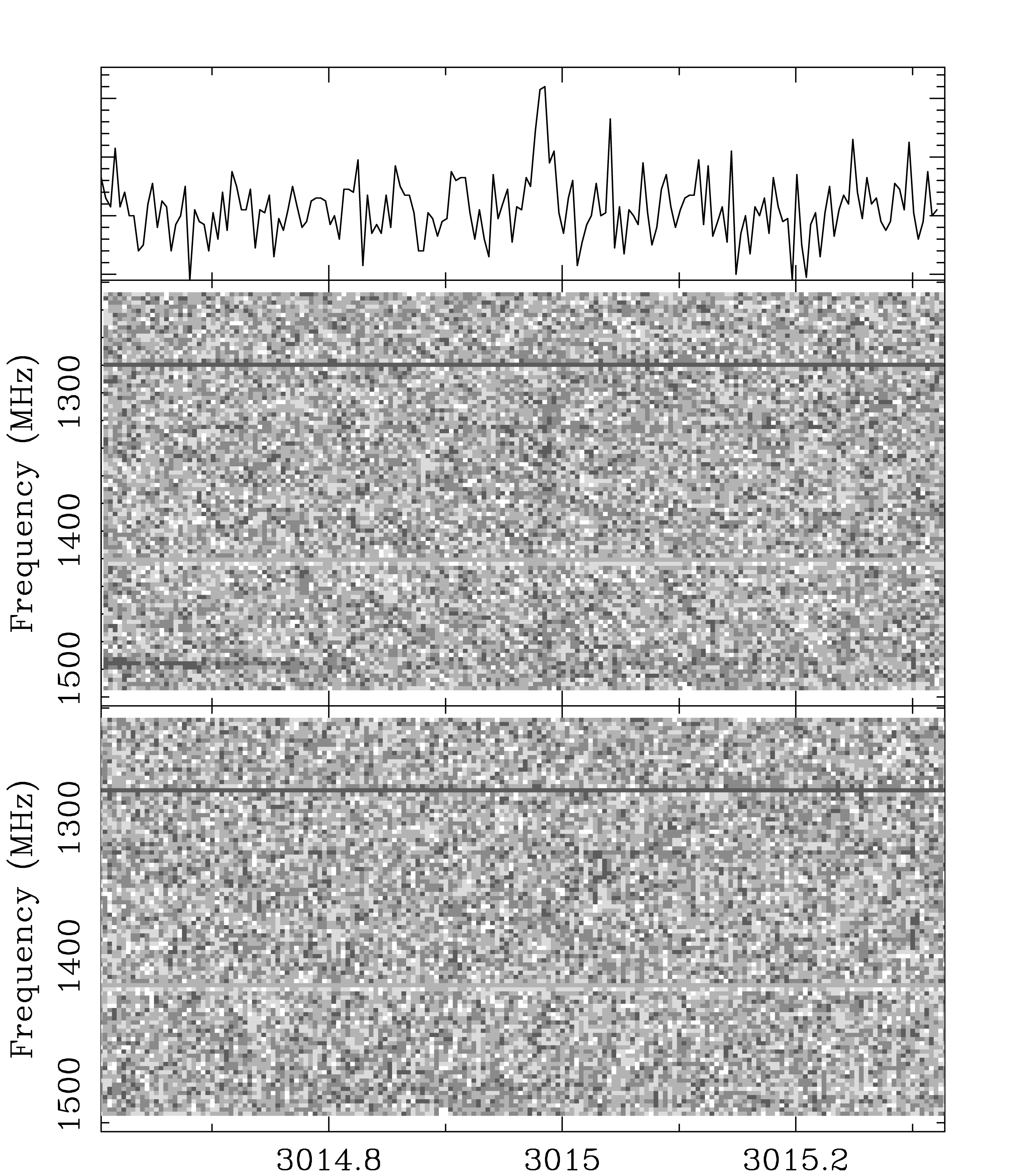} & 
    \includegraphics[width=5.5cm]{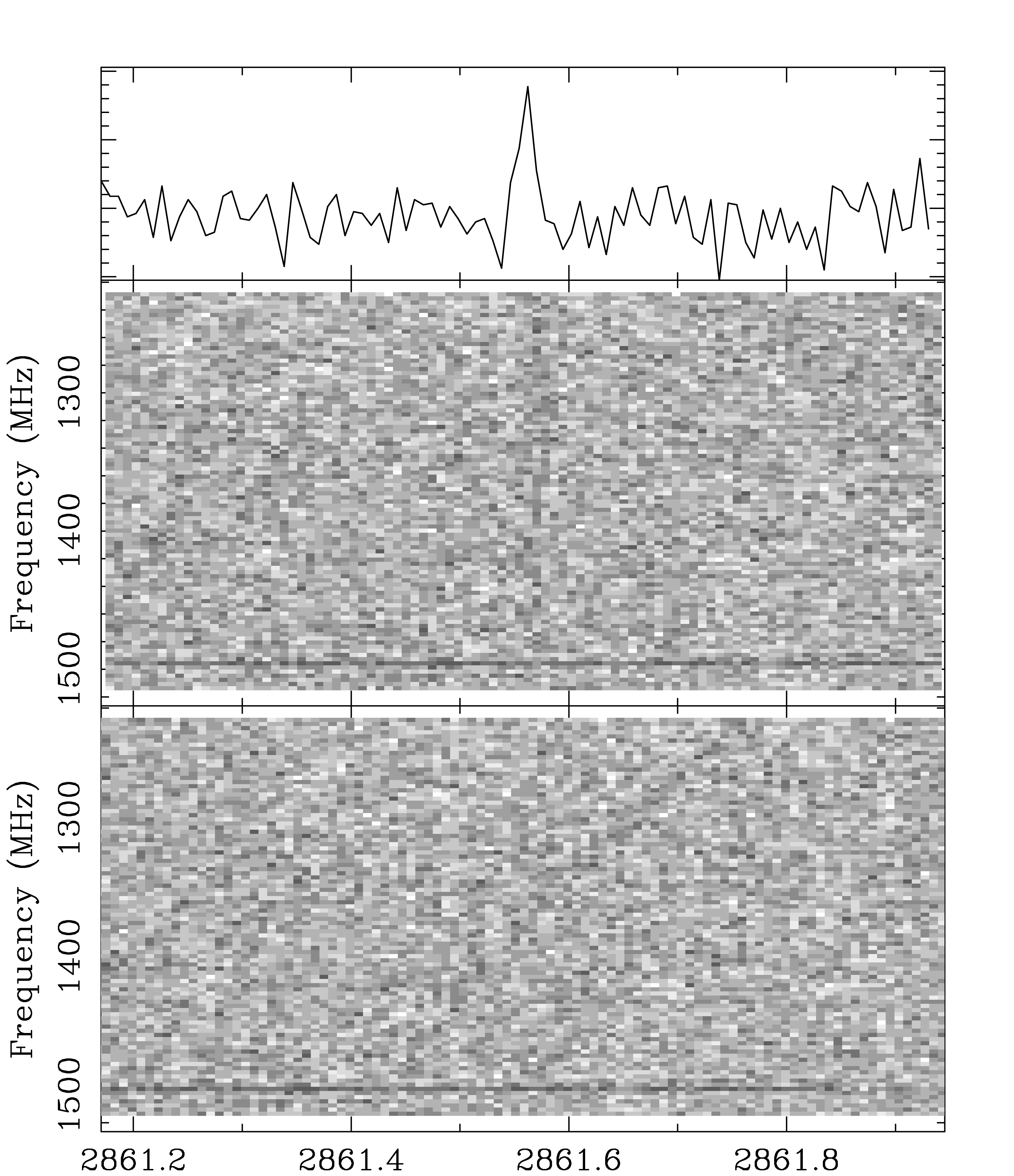}\\
    \end{tabular}
    \caption{Pulse profiles of six example candidate FRBs detected in Parkes data archive. For each sub-figure, the bottom panel shows the time-frequency plane, the central panel shows the event after being de-dispersed at the optimised DMs, while the top panel is the integrated pulse profile using an arbitrary flux density scale. From left to right, the S/N of these candidate FRBs are 7.1, 7.2, 7.5, 7.5, 8.0, 8.1, respectively.} 
    \end{center}
    \label{figure:frb_spc}
\end{figure*}

In our previous data reduction pipeline, we identified the observations in which burst events occurred.  For each event we are able to plot dynamic spectra around each event. Examples are shown in Figure \ref{figure:frb_spc}.

\begin{figure*}
	\includegraphics[width=\textwidth]{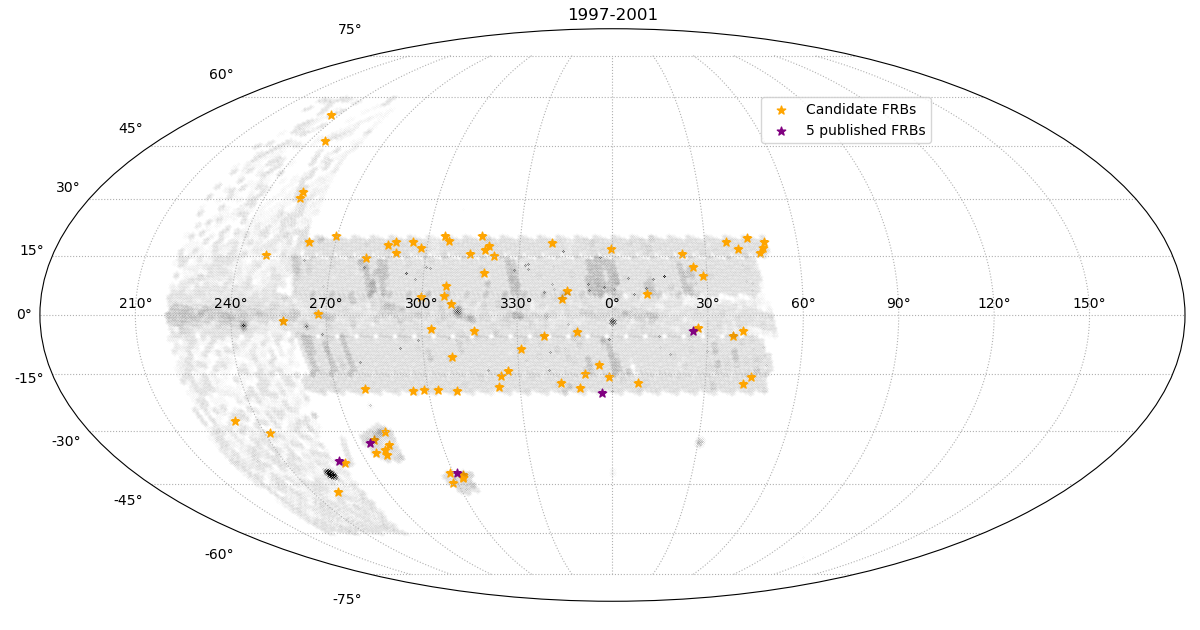}
    \caption{
    The sky map of the Parkes data used in this work in Galactic coordinates. 
    The orange stars show the distribution of 81 candidate FRBs, and the shadow zone shows the survey regions from 1997 to 2001. The 5 purple points are published FRB detections from Parkes in 2001.}
    \label{fig:3}
\end{figure*}

The data sets used in creating the PTD contain all of the observations during the first four years (from 1997 to 2001) of operation of the Parkes Multibeam receiver system. The central frequency, bandwidth and number of channels for these observations were 1374~MHz, 288~MHz, and 96 channels, respectively.   The sky coverage of these observations are shown in Figure \ref{fig:3}. 
Candidates with adjacent DMs, and overlapping start and end times, often derive from the same wide-profile signal. Therefore, the candidates that lie within the start and end time of a given event are grouped and defined as a file segment~\citep[see more details on file segments in][]{Zhang_2020}.
Our classification pipeline makes use of \emph{\sc sqlite3}~\footnote{\url{https://www.sqlite.org/}} to automatically select a candidate with the largest S/N from each file segment. We then produce time-frequency plots as shown in Figure \ref{figure:frb_spc}.

To optimise the processing speed, we extracted  dynamic spectra images without applying any de-dispersion. The extracted image size is $850\times680$ pixels. We chose bicubic interpolation to downsize the figure into $256\times256$ pixels.  An eighteen-layer residual neural network (ResNet-18) was then used to process the $256\times256$ pixels figures. 
We chose \emph{\sc pytorch}~\footnote{\url{https://pytorch.org}} to implement the ResNet-18 with $\rm pretrained=False$. The first layer of the ResNet-18 contains 64 convolution kernels with a kernel size of $k^2=7\times7$, a stride of $s=2$ and a padding of $p=3$ \citep[see][for more details]{heresnet}. The output size of the vectors for a particular dimension can be calculated as
\begin{equation}
    y_{\rm size}=\left|\frac{x_{\rm size}+2 \times p-k}{s}\right|+1,
	\label{eq:size}
\end{equation}
where 
$|...|$ means the quotient operation. Our input figure, which contains $x_{\rm size}^2=256\times256$ pixels becomes 64 smaller vectors with a size of $y_{\rm size}^2=128\times128$ pixels.
Following \citet{heresnet}, we next adopt a maximum pooling process, which can downsize the vectors to the size of $64\times64$ pixels. 
The second to seventeenth layers contain 16 convolutional networks with a kernel size of $k^2=3\times3$ and eight times of operating with $W_sx$. These layers convert the vectors into 512 vectors with a size of $y_{\rm size}^2=8\times8$ pixels. The eighteenth layer is a full connection layer which combines the small vectors together and generates the final result which indicates whether the event is more likely to be an FRB or RFI.

We manually built a training set of 3,000 images to train the machine learning algorithm. These figures included 1,500 labelled examples of RFI, 1,500 single pulses from pulsars, and five previously published FRBs. 
The 1,500 examples of RFI contain several types including narrow-band and broad-band RFI, periodic RFI, and RFI that saturated the image. 
After 100 training epochs, the training accuracy of our ResNet-18 implementation was 98.6\%, i.e., 98.6\% of figures in the training set are correctly identified. We also built a test set with 500 labelled examples of RFI and 500 single pulses from pulsars which are not part of the training set. For these, the test accuracy was 95.9\%.

We applied the trained ML system to 808,330 events. The algorithm identified 335,877 events as RFI and the remaining 472,453 events as single pulses.
The PTD contains information on all the known pulsars. We compared the sky position for the observation and the candidate DMs with known sources to identify that 451,609 of the single pulse candidates are from known pulsars. 

After excluding these pulsar events, we obtained a relatively small sample of 20,644 events. As we aimed to search for more candidate FRBs, we used the Galactic coordinates for the telescope pointing directions to determine $\dmmw$~(the DM contributed by our Galaxy) based on the YMW16 electron density model~\citep{Yao2017}. 
Only events where DM$_{\rm excess}=$DM$_{\rm obs}-$DM$_{\rm MW}>0$ were selected as possible candidate FRBs. Through this procedure, the sample was reduced to 6,409 events. Finally, we visually inspected those figures to study whether the dynamic spectra for these signals conformed to the dispersion relation and if the event was detected was in three or fewer adjacent beams. Nine ``perytons'' were identified, which were detected in each of the multibeam beams simultaneously and have been identified as being generated when a microwave oven door is opened prematurely on-site \citep[]{perytons}. 


\section{Results}


\label{sec:result}

After visual inspection, we obtained 81 new candidate FRBs as well as five previously published FRBs. 
The properties of these candidate FRBs are listed in Table \ref{table:properties}. 
They were all detected in one beam of the Parkes multibeam system with ${\rm DM}_{\rm excess}>0$. 
Three have S/N $\ge 8.0$, 28  with S/N $\sim7.5-8.0$, and 51 with S/N $\sim7.0-7.5$.
Frequency-time plots and dedispersed pulse profiles are presented in Figure~\ref{figure:frb_spc} for six of the candidate FRBs. In these six, we have chosen two relatively low S/N candidate FRBs ($\sim$7), two relatively high S/N candidate FRBs ($\sim$8), and two candidate FRBs whose S/N are around 7.5. The properties and pulse profiles for all 81 candidate FRBs are available from \url{https://astroyx.github.io/}.
Figure \ref{fig:3} shows the location distribution of the 81 candidate FRBs (marked as orange stars) and the five  previously-published FRBs (marked as purple stars).

\begin{figure*}
	\includegraphics[width=0.98\textwidth]{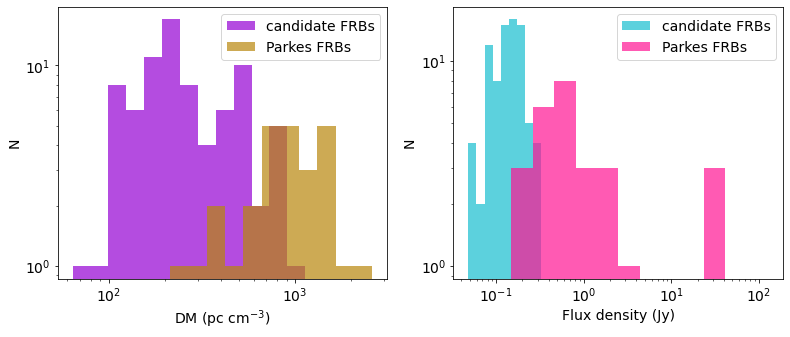}
    \caption{The histograms of the DM and peak flux density distributions of the 81 candidate FRBs versus 28 published Parkes FRBs. }
    \label{fig:4}
\end{figure*}

The peak flux density for any event can be estimated from
\begin{equation}
    S_{\rm peak}=\frac{\sigma \  S/N \ T_{\rm sys}}{G \sqrt{\Delta \nu  N_p t_{\rm wid}}},
	\label{eq:slim}
\end{equation}
where $\sigma=1.5$ is a loss factor \citep{Manchester2001}, $T_{\rm sys}$ is the system temperature, $G$ is the telescope antenna gain\footnote{\url{https://www.parkes.atnf.csiro.au/observing/documentation/users\_guide/html/}}, $\Delta \nu$ is the observing bandwidth, $N_p$ is the number of polarisation channels, and $t_{\rm wid}$ is the duration of the burst.
Histograms of the candidate DM and peak flux density values are shown in Figure \ref{fig:4}. The candidate DMs range from 64 to 1,131~pc\,cm$^{-3}$ whilst the peak flux densities of candidate FRBs range from 0.05 to 0.33~Jy. 

\begin{figure*}
\includegraphics[width=0.98\textwidth]{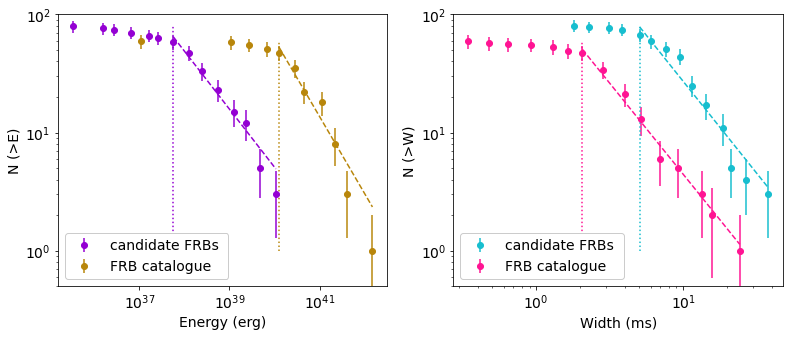}
    \caption{The cumulative distribution of the isotropic energy and width of the candidate FRBs versus the FRB catalogue. Power-law distribution have been fitted to the tails. The vertical lines represent cut-off energies and cut-off widths. Form left to right, the values of vertical lines are  $5.6\times10^{37}$\,erg, $1.3\times10^{40}$\,erg, 2.1\,ms and 5.1\,ms respectively.}
    \label{fig:5}
\end{figure*}

%
The distance and redshift of each candidate FRB was estimated using the YMW16 model with ${\rm DM}_{\rm host}$=0.
We then also estimate the isotropic equivalent peak luminosity $(L_{\rm peak})$ and corresponding isotropic energy $(E_{\rm iso})$\footnote{Note that the estimated distance and the corresponding redshift given by the YMW16 model have significant uncertainties from each of the DM contributions, with the inter-galactic and FRB host contributions being the largest, and hence the corresponding energy $(E_{\rm iso})$ also has a large uncertainty. } for these candidate FRBs. 
We find that they vary over several orders of magnitudes, that is, $L_{\rm peak}\sim 10^{38}$\,erg\,s$^{-1}-10^{42}$\,erg\,s$^{-1}$, and $E_{\rm iso}\sim10^{35}$\,erg$-10^{40}$\,erg.
Figure \ref{fig:5} shows the cumulative distribution of isotropic energy and width of the candidate FRBs. We compare these distributions with those from 59 events detected at a centre frequency around 1.3\,GHz in the FRB catalogue\footnote{
The catalogue contains the population of FRBs published up to July 2020, available from \url{http://www.frbcat.org}~\citep{petroff_frbcat}. In this work, 28, 26, 1, 1, and 1 FRBs detected by Parkes, ASKAP, FAST, Apertif, VLA telescopes were used.}. Dotted oblique lines show the power-law fitting of these distribution, and dotted vertical lines show the cut-off energy $E_{\rm cut}$. 
$E_{\rm cut}$ is the first point used for fitting and is chosen by comparing the edge between two distributions of a sample set.
We find that the power-law indices of these distributions are $\alpha_1=-1.5\pm0.1$ and $\alpha_2=-1.7\pm0.1$ for the energy distributions of candidate FRBs and FRB catalogue respectively, $\beta_1=-2.6\pm0.1$ and $\beta_2=-2.5\pm0.1$ for the width distributions of candidate FRBs and FRB catalogue respectively.
The cut-off energies are $E_{\rm cut,1}=5.6\times10^{37}$\,erg and $E_{\rm cut,2}=1.3\times10^{40}$\,erg for the candidate FRBs and the FRB catalogue respectively.

\section{Discussion}
\label{sec:Dis}

The primary goal of this work was to develop a technique that could classify events drawn from a large sample. Applying the ResNet-18 ML model to the PTD led to 81 relatively strong candidate FRBs with only minor human input.   Being able to cut down candidate numbers automatically is becoming more and more essential. For example, FAST \citep{fastcandidate} and the Square Kilometre Array \citep{Dewdney2009} are likely to obtain billions of candidates. 

\begin{figure}
\begin{center}
\includegraphics[width=0.45\textwidth]{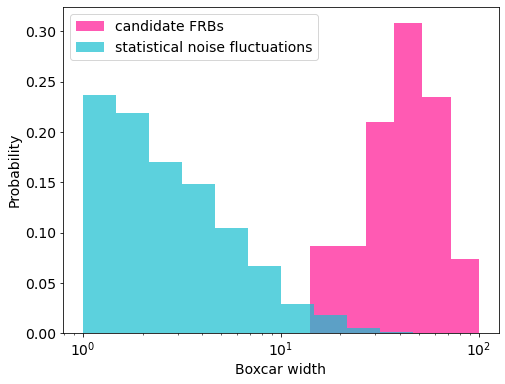}
    \caption{The histograms of the boxcar widths of the 81 candidate FRBs and candidates detected in the simulated data sets containing only normally distributed noise.}
    \label{fig:6}
\end{center}
\end{figure}

To analyse how many of these 81 candidate FRBs could come from statistical noise fluctuations, we simulated ten 1-hour Parkes data sets containing only normally distributed noise. The central frequency, bandwidth, number of channels and sample time used in our simulations were 1374~MHz, 288~MHz, 96 channels and 1~ms, respectively. We then processed these data sets in the same way as was used in the PTD~\footnote{Based on the duration and the corresponding DM$_{\rm MW}$ of the observations in the PTD, an average DM$_{\rm MW}$ of 442.1 pc$\cdot$cm$^{-3}$ is obtained. Only events with DMs above this value were selected.}. Processing the ten simulated data sets and taking the average candidate count for each 1-hour observation gave an average of 489.4 candidates with S/N $\ge 5.0$, 45.4 with S/N $\ge 5.5$, and 2.1 with S/N $\ge 6.0$. This result is consistent with a normal distribution,\footnote{Depending on the datasets and our processing pipelines, the number of statistical trials in a 1-hr observation is $\sim 2.33 \times 10^{9}$. } therefore, we speculate that $\sim 2.98\times 10^{-3}$, $\sim 7.42 \times 10^{-5}$ and $\sim 1.55 \times 10^{-6}$ such events would be detected with S/N $\ge 7.0$, S/N $\ge 7.5$ and S/N $\ge 8.0$, respectively. However, as shown in Figure~\ref{fig:6}, the smallest boxcar width\footnote{\emph{\sc PRESTO} convolves the dedispersed time series data with different boxcar widths. When an event is detected at multiple boxcar widths, it filters the boxcar width list to report only the width with the best S/N.} of our 81 candidate FRBs is 14, and only $\sim 10.1\%$ of the candidates in the pure noise data sets were detected with boxcar widths $\ge 14$. As the total integration time of the PTD is 38,190~hr, about 11.5 with S/N $\ge 7.0$, 0.3 with S/N $\ge 7.5$ and 0.006 with S/N $\ge 8.0$ of our 81 candidate FRBs could be result from statistical noise fluctuations. This implies that most of our candidate FRBs, especially those with S/N $\ge 7.5$, are not from background, Gaussian noise. They may therefore be real, or may be from RFI that mimics a high DM candidate. 
However, we note that the ML algorithm is trained to distinguish FRBs from RFI. RFI , like ``perytons'', could be indistinguishable for our ML algorithm, but would be eliminated by the number-of-beams criterion.

The range of the peak flux densities of our candidate FRBs is relatively small compared with the Parkes FRBs, which is consistent with our search using a lower S/N threshold than in previous work.
The mean DM for these candidates is smaller than the mean DM of previously published FRBs detected using the Parkes telescope. We note that the observations in the PTD are less sensitive to high-DM FRB events, because of the use of relatively wide channel widths (i.e. 3\,MHz) in these early observations. 
A wide channel width would also result in large DM-smearing~\citep{dmsmearing}. For instance, the intra-channel DM smearing across a 3\,MHz channel is $\sim$ 3.1\,ms at 1374\,MHz for a signal with a DM of the mean DM of our 81 candidate FRBs of 319.5\,cm$^{-3}$ pc. We noticed that the observed width of each of our candidate FRB is larger than its intra-channel DM smearing broadening width.

Comparing with all the events in the FRB Online Catalogue, the range of most of the candidate FRBs are of peak luminosity and energy consistent with these published FRBs. While for some candidate FRBs, the luminosity and energy are much lower than the extra-galactic FRBs, but consistent with the Galactic one \citep{STARE2}. 
As shown in Figure \ref{fig:5}, the cut-off energy of our candidate FRBs sample is $E_{\rm cut,1}=5.6\times10^{37}$\,erg, much smaller than that ($E_{\rm cut,2}=1,3\times10^{40}$\,erg) of the events of the current FRB catalogue. 
The difference in the cut-off energies could be caused by (1) the S/N of all our candidate FRBs being relatively low, so their flux densities are small; (2) the PTD observation is less sensitive to high-DM FRB events indicating larger distances; (3) almost half of the used FRBs in the catalogue were discovered with telescopes less sensitive to Parkes radio telescope and (4) we have included all the events with ${\rm DM}_{\rm excess}>0$ therefore some candidate FRBs may have very small inferred distances.
The cut-off width of our candidate FRBs sample is $\sim 3\,$ms smaller than that of the events of the current FRB catalogue, which is consistent with the intra-channel DM smearing of the PTD data. 
We find the power-law index of the energy and width distributions of the candidates are consistent with those of the FRBs in the FRB catalogue, which provides evidence that these candidates could be real FRBs.

These candidates can be used to guide follow-up observations. In particular for the low-DM-excess events, they may be FRB events from nearby galaxies, and thus would be ideal for searching for electromagnetic counterparts, such as X-ray bursts \citep{HXMT2020,INTEGRAL2020,Konus2020,Geng20,Dai2020ApJ}, multi-wavelength nebulae for repeating FRBs \citep[e.g.][]{Chatterjee17,Wang2020a}, and possible multi-wavelength afterglows for individual events of FRBs \citep{Wang2020a}. 

Some of these low-DM-excess events could also come from Galactic pulsars or from rotating radio transients (RRATs). A detection from a source located in the Large Magellanic Cloud (LMC) or Small Magellanic Cloud (SMC), whose DM-excess could generated by the surrounding nebula, would also allow us to obtain a higher FRB event rate compared with other surveys. 
Taking into account the survey integration time of 38,190 hours, 86 events (81 candidate FRBs and 5 published FRBs) in a total of four years of observing correspond to an all-sky detectable event rate of $5.2^{+1.0}_{-0.9}\times 10^4~\rm events~d^{-1}$.
Table \ref{table:rate} lists the on-sky integration time in three latitude ranges, and the corresponding event rate at the 95 percent
confidence level. The event rates obtained from this work are approximately ten times larger than the estimates of Parkes HTRU survey~\citep{Bhandari2017} and are approximately half of the estimate from the FAST CRAFTS survey~\citep{Niu2021}. 
Our event rate estimate could be over-estimated as we used a lower flux density threshold in the FRB search and included all of the events (most of which are close to the detection threshold).
We also note that our relatively high event rate estimates may be partly caused by some of the signals being Galactic sources (e.g., from pulsars or RRATs) and some may simply be coincidental RFI. 


\begin{table}
\caption{The on-sky integration time, in three latitude ranges, for observations from 1997 to 2001. The all-sky event rates are shown with 95 percent confidence level.\label{table:rate}}
\renewcommand\arraystretch{1.5}    
\begin{center}
\begin{tabular}{@{\extracolsep{\fill}}c|c|c|c}
\hline

\multicolumn{1}{c|}{Galactic latitude (deg)}  &     Total time (h)        &     $N_{\rm events}$  	 &     	$ R_{\rm FRB}~\rm (events~d^{-1})$        \\\hline
$0^\circ \le \left|b\right| \leq 19.5^\circ$    &   29,938  &   58   &    $4.5^{+1.1}_{-0.9}\times 10^4$       \\\hline 
$19.5^\circ < \left|b\right| \leq 42^\circ$    &   5,799  &   21   &    $8.4^{+3.7}_{-2.8}\times 10^4$       \\\hline 
$42^\circ < \left|b\right| \leq 90^\circ$    &   2,453  &   7   &    $6.6^{+5.9}_{-3.5}\times 10^4$        \\\hline
\end{tabular}
\end{center}
\end{table}

 Searching for weak FRB events from large data sets can help study the relationship between ``repeating'' and ``non-repeating'' bursts. We notice the  cut-off energy and power-law index of our candidate FRBs sample are $\alpha_{1}=-1.5\pm0.1$ and $E_{\rm cut,1}=5.6\times10^{37}$\,erg, very close to the results of a
sample of low-energy bursts from the repeating FRB~121102 [i.e. $\alpha=-1.8\pm0.3$ and $E_{\rm cut}=2\times10^{37}$\,erg, \citep{Gourdji2019}]. However, the cut-off energy of FRB~121102 of~\citet{Gourdji2019} was recognized as due to the completeness limit of the telescope. Note that most of our lower energy candidate FRBs have DM$_{\rm excess}$ < 50\% DM$_{\rm MW}$ and our estimated energies $(E_{\rm iso})$ also have large uncertainties, it is hard to make any convincing conclusion or indication with our cut-off energy.
The similarity is not yet highly significant and more reliable samples are necessary to make any prediction. 
The huge number of relatively weak FRBs is also crucial in determining the luminosity function of FRBs, and hence the accurate event rate. While different physical models predict different event rates, in the future, searching for weak signals will be required to constrain these models.
This is the first attempt to apply a ML algorithm to the first version of the PTD and we have identified a large number of candidates for follow-up observations.  In the near future we will continue to enlarge the PTD with more recent observations and also will carry out a more in-depth study of the various ML algorithms that are available.  In particular we will inject FRB-like signals into the data sets allowing us to form a quantifiable measure of the effectiveness of these algorithms.


\section*{Acknowledgements}
The Parkes radio telescope (``Murriyang'') is part of the Australia Telescope National Facility which is funded by the Australian Government for operation as a National Facility managed by CSIRO. This paper includes archived data obtained through the CSIRO Data Access Portal (\url{https://data.csiro.au}). This work was supported by ACAMAR Postdoctoral Fellow, the National Natural Science Foundation of China (Grant No. 11725314, 12041306, 11903019), China Postdoctoral Science Foundation (Grant No. 2020M681758), the Guizhou Provincial Science and Technology Foundation (Grant No. [2020]1Y019). Parts of this research were supported by the Australian Research Council Centre of Excellence for All Sky Astrophysics in 3 Dimensions (ASTRO 3D), through project number CE170100013. SD is the recipient of an Australian Research Council Discovery Early Career Award (project DE210101738) funded by the Australian Government.

\section*{DATA AVAILABILITY STATEMENTS}
The data underlying this article include (1) the Parkes archived observation which are available for download from CSIRO Data Access Portal (\url{https://data.csiro.au}); (2) the Parkes Transient Database which is available from \url{https://data.csiro.au/dap/landingpage?pid=csiro:42640}; (3) the properties and pulse profiles for all 81 candidate FRBs which are available from \url{https://astroyx.github.io/}.

\begin{table*}
  \begin{scriptsize}
  \caption{The properties of 81 new candidate FRBs detected in this work. The five previously published FRBs are not listed in this table.\label{table:properties}}
  \renewcommand\arraystretch{0.95}
  \setlength{\tabcolsep}{2.0mm}  
  \begin{center}
  \begin{tabular}{@{\extracolsep{\fill}}c|c|c|c|c|c|c|c|c|c}
  \hline
  
\multicolumn{1}{c|}{Filename}  &    RA      &  Dec        &     Galactic longitude   	 &     	Galactic latitude        &    DM$_{\rm obs}$        &     DM$_{\rm MW}$       &      S/N       &       	Width         &           Beam number      \\
 \multicolumn{1}{c|}{}            &            &             &                       	     &                                 &     (pc$\cdot$cm$^{-3}$)       &    (pc$\cdot$cm$^{-3}$)        &             &       	(ms)         &                  \\
 \hline
  SMC012\_01281.sf    &   00:45:13.9  &   -74:49:04.8   &    303.4812    &    -42.3036    &    259.9    &    32.0    &    7.2    &    9.8    &    8    \\ 
  SMC012\_01221.sf    &   00:49:18.1  &   -74:24:31.6   &    303.1274    &    -42.7187    &    1013.0    &    31.7    &    7.3    &    33.7    &    2    \\ 
  SMC012\_01231.sf    &   00:53:05.2  &   -73:59:53.0   &    302.7763    &    -43.1297    &    382.5    &    31.3    &    7.4    &    11.2    &    3    \\ 
  SMC014\_059C1.sf    &   01:35:18.0  &   -72:01:17.6   &    298.1937    &    -44.7010    &    528.9    &    30.1    &    8.0    &    17.7    &    12    \\ 
  SMC020\_00171.sf    &   01:39:32.2  &   -75:01:24.2   &    298.7958    &    -41.7163    &    422.9    &    32.9    &    7.4    &    11.3    &    7    \\ 
  PH0012\_05341.sf    &   04:09:10.3  &   -44:29:06.3   &    250.1791    &    -47.0918    &    214.8    &    29.0    &    7.2    &    7.5    &    4    \\ 
  SMC014\_038D1.sf    &   04:46:50.2  &   -67:15:29.2   &    278.5401    &    -36.7630    &    853.0    &    47.2    &    7.5    &    40.3    &    13    \\ 
  SM0005\_03551.sf    &   04:53:38.2  &   -53:49:47.5   &    261.6881    &    -38.7817    &    211.4    &    36.2    &    7.6    &    10.3    &    5    \\ 
  SMC019\_01391.sf    &   05:00:04.5  &   -67:55:11.4   &    278.8776    &    -35.3649    &    592.3    &    50.8    &    7.6    &    26.7    &    9    \\ 
  SMC016\_00181.sf    &   05:00:40.0  &   -64:57:46.2   &    275.3282    &    -36.0277    &    368.6    &    48.0    &    8.1    &    18.4    &    8    \\ 
  SMC014\_006A1.sf    &   05:09:55.0  &   -69:59:49.7   &    281.0488    &    -33.9843    &    206.6    &    54.6    &    7.5    &    20.2    &    10    \\ 
  SMC017\_005B1.sf    &   05:31:52.8  &   -66:44:56.6   &    276.7940    &    -32.5800    &    508.9    &    58.3    &    7.5    &    21.0    &    11    \\ 
  PH0015\_07571.sf    &   05:34:42.2  &   -37:02:41.7   &    241.8570    &    -30.5389    &    478.5    &    45.8    &    7.7    &    13.8    &    7    \\ 
  PH0038\_10881.sf    &   05:38:49.3  &   -28:10:59.7   &    232.3412    &    -27.3730    &    114.2    &    54.8    &    7.8    &    18.3    &    8    \\ 
  SMC017\_073C1.sf    &   05:51:45.0  &   -71:14:01.4   &    281.8149    &    -30.3626    &    1131.4    &    66.4    &    7.2    &    39.6    &    12    \\ 
  BJ0006\_063D1.sf    &   07:54:30.5  &   -67:04:34.8   &    279.5738    &    -19.0124    &    165.1    &    154.6    &    7.3    &    3.9    &    13    \\ 
  PMM001\_05071.sf    &   08:19:18.6  &   -38:38:28.9   &    256.4980    &    -1.4149    &    579.9    &    476.4    &    7.3    &    13.3    &    7    \\ 
  PH0030\_003A1.sf    &   08:59:42.6  &   -22:19:07.8   &    248.5657    &    15.3049    &    479.7    &    75.7    &    7.7    &    9.9    &    10    \\ 
  PM0004\_00651.sf    &   09:02:03.8  &   -46:14:24.0   &    267.3188    &    0.1615    &    485.1    &    449.4    &    7.4    &    8.5    &    5    \\ 
  BJ0010\_035B1.sf    &   09:46:17.2  &   -79:27:25.8   &    294.8606    &    -19.6185    &    210.2    &    108.3    &    7.5    &    6.6    &    11    \\ 
  BJ0005\_006D1.sf    &   09:47:36.9  &   -28:51:31.9   &    261.2568    &    18.7724    &    189.7    &    155.1    &    7.5    &    6.9    &    13    \\ 
  PH0037\_08251.sf    &   09:56:44.9  &   -14:50:39.7   &    252.2568    &    30.3595    &    64.0    &    41.8    &    7.7    &    2.3    &    5    \\ 
  PH0026\_07231.sf    &   10:01:49.3  &   -13:54:00.7   &    252.4855    &    31.8966    &    163.5    &    40.0    &    7.5    &    5.5    &    3    \\ 
  BJ0014\_04851.sf    &   10:20:02.9  &   -32:24:18.7   &    269.3747    &    20.4247    &    126.8    &    84.6    &    7.2    &    3.7    &    5    \\ 
  PH0042\_07061.sf    &   10:30:27.2  &   +00:25:28.7   &    246.2259    &    46.3390    &    114.8    &    28.1    &    7.2    &    6.6    &    6    \\ 
  PH0034\_10311.sf    &   10:39:41.3  &   +09:16:00.2   &    236.5260    &    54.1560    &    646.3    &    24.4    &    7.2    &    11.8    &    1    \\ 
  BJ0014\_05881.sf    &   10:52:59.7  &   -43:03:39.4   &    281.0160    &    14.7140    &    171.9    &    148.1    &    7.2    &    3.6    &    8    \\ 
  BJ0016\_04661.sf    &   10:58:43.4  &   -81:05:48.3   &    298.4931    &    -19.2126    &    225.9    &    104.0    &    7.3    &    6.7    &    6    \\ 
  BJ0024\_056D1.sf    &   11:29:53.8  &   -42:24:36.1   &    287.2459    &    17.9645    &    225.4    &    124.5    &    7.4    &    6.6    &    13    \\ 
  BJ0005\_03861.sf    &   11:41:06.5  &   -45:10:08.9   &    290.1351    &    15.9479    &    294.9    &    144.9    &    7.6    &    9.7    &    6    \\ 
  BJ0005\_032B1.sf    &   11:43:21.1  &   -42:29:47.4   &    289.7661    &    18.6279    &    216.8    &    115.4    &    7.7    &    8.8    &    11    \\ 
  BJ0010\_01281.sf    &   12:11:13.8  &   -43:30:58.2   &    295.2500    &    18.7550    &    751.2    &    105.0    &    8.1    &    10.6    &    8    \\ 
  BJ0005\_05321.sf    &   12:25:51.7  &   -45:27:48.0   &    298.2425    &    17.1754    &    193.6    &    112.6    &    7.5    &    3.7    &    2    \\ 
  PM0112\_05851.sf    &   12:27:03.6  &   -58:04:15.7   &    299.7024    &    4.6478    &    513.7    &    330.4    &    7.2    &    18.3    &    5    \\ 
  PM0110\_04231.sf    &   12:50:21.0  &   -66:29:33.3   &    302.8232    &    -3.6211    &    829.6    &    539.0    &    7.5    &    10.5    &    3    \\ 
  BJ0003\_033C1.sf    &   12:58:20.0  &   -82:06:12.0   &    303.1827    &    -19.2349    &    103.6    &    96.2    &    7.6    &    1.8    &    12    \\ 
  BJ0017\_00831.sf    &   13:03:53.7  &   -42:35:23.9   &    305.3747    &    20.2227    &    104.4    &    83.8    &    7.3    &    1.8    &    3    \\ 
  BJ0017\_01381.sf    &   13:11:18.8  &   -43:45:03.0   &    306.7254    &    18.9745    &    169.0    &    89.7    &    7.3    &    3.8    &    8    \\ 
  PM0108\_024A1.sf    &   13:22:07.5  &   -57:44:24.8   &    307.0330    &    4.8870    &    440.3    &    283.2    &    7.5    &    10.3    &    10    \\ 
  SW0007\_08541.sf    &   13:23:24.9  &   -55:17:47.2   &    307.5104    &    7.2908    &    257.9    &    214.8    &    7.5    &    14.1    &    4    \\ 
  PM0080\_03581.sf    &   13:42:01.5  &   -59:25:54.4   &    309.3325    &    2.8099    &    905.5    &    468.8    &    7.7    &    8.8    &    8    \\ 
  BJ0021\_067D1.sf    &   13:54:02.1  &   -45:58:54.1   &    314.1499    &    15.5221    &    256.5    &    106.1    &    7.1    &    7.5    &    13    \\ 
  BJ0001\_10081.sf    &   14:02:32.5  &   -40:39:59.3   &    317.2193    &    20.2243    &    131.7    &    78.7    &    7.3    &    4.9    &    8    \\ 
  SW0004\_02281.sf    &   14:12:21.7  &   -72:36:22.2   &    308.9706    &    -10.6900    &    410.0    &    175.6    &    7.7    &    10.1    &    8    \\ 
  BJ0001\_09211.sf    &   14:17:04.0  &   -43:37:31.0   &    318.9330    &    16.5700    &    102.1    &    96.9    &    7.2    &    4.5    &    1    \\ 
  BJ0002\_1\_00211.sf    &   14:19:21.5  &   -42:10:49.7   &    319.8660    &    17.7820    &    107.9    &    89.8    &    7.3    &    5.3    &    1    \\ 
  SW0004\_104D1.sf    &   14:31:12.9  &   -49:00:21.6   &    319.2830    &    10.6721    &    206.6    &    147.1    &    7.4    &    7.8    &    13    \\ 
  BJ0002\_1\_013C1.sf    &   14:34:29.3  &   -43:56:08.8   &    321.8500    &    15.1180    &    202.1    &    105.1    &    7.5    &    10.4    &    12    \\ 
  PM0120\_00791.sf    &   14:56:40.9  &   -63:35:14.4   &    316.3317    &    -4.0225    &    473.3    &    472.6    &    7.9    &    11.1    &    9    \\ 
  BJ0017\_058C1.sf    &   15:26:52.9  &   -80:18:18.3   &    309.3652    &    -19.4369    &    172.4    &    88.7    &    7.2    &    4.6    &    12    \\ 
  BJ0016\_068C1.sf    &   15:41:23.5  &   -31:43:25.1   &    340.2388    &    18.5704    &    141.8    &    82.5    &    7.2    &    9.2    &    12    \\ 
  SW0007\_02341.sf    &   16:41:47.5  &   -37:06:29.4   &    345.5438    &    6.0461    &    581.6    &    240.0    &    7.5    &    8.4    &    4    \\ 
  BJ0002\_2\_02611.sf    &   16:43:14.7  &   -19:35:59.1   &    359.5330    &    16.9740    &    112.8    &    91.1    &    7.5    &    5.4    &    1    \\ 
  PM0130\_023D1.sf    &   16:44:40.0  &   -39:29:44.2   &    344.0986    &    4.0596    &    382.9    &    326.6    &    7.4    &    8.3    &    13    \\ 
  SM0025\_010D1.sf    &   16:52:51.2  &   -57:45:02.2   &    330.9088    &    -8.6695    &    378.4    &    187.5    &    7.3    &    13.0    &    13    \\ 
  PM0131\_00611.sf    &   17:04:59.7  &   -49:51:28.1   &    338.3000    &    -5.2540    &    730.2    &    305.5    &    8.0    &    19.9    &    1    \\ 
  BJ0003\_089A1.sf    &   17:07:24.4  &   -67:04:11.8   &    324.2434    &    -15.5397    &    364.0    &    104.4    &    7.9    &    10.4    &    10    \\
    BJ0004\_01181.sf    &   17:09:25.1  &   -64:33:42.0   &    326.5160    &    -14.3100    &    128.6    &    112.8    &    7.2    &    3.0    &    8    \\ 
  BJ0003\_07631.sf    &   17:28:16.9  &   -69:20:20.7   &    323.2714    &    -18.3730    &    139.5    &    87.8    &    7.1    &    3.8    &    3    \\
  PM0123\_04141.sf    &   17:34:29.5  &   -40:33:34.5   &    348.9261    &    -4.2281    &    374.3    &    344.4    &    7.8    &    8.8    &    4    \\ 
  BJ0026\_00271.sf    &   17:36:46.8  &   -02:05:32.7   &    22.2654    &    15.5440    &    197.4    &    115.3    &    7.3    &    5.0    &    7    \\
  PM0141\_00181.sf    &   17:49:34.5  &   -16:47:60.0   &    10.8860    &    5.4744    &    319.5    &    273.5    &    7.3    &    5.7    &    8    \\ 
  BJ0020\_09651.sf    &   17:50:37.4  &   +11:54:00.5   &    36.9610    &    18.8087    &    182.3    &    73.3    &    7.8    &    5.4    &    5    \\ 
  SW0005\_089A1.sf    &   17:54:44.3  &   +00:38:20.0   &    25.7841    &    12.2891    &    210.2    &    123.2    &    7.4    &    15.7    &    10    \\ 
  BJ0022\_02641.sf    &   17:58:05.3  &   +18:26:49.7   &    44.0764    &    19.8187    &    226.7    &    67.0    &    7.3    &    10.1    &    4    \\ 
  BJ0023\_05871.sf    &   18:03:29.8  &   +14:17:36.3   &    40.6204    &    16.9580    &    212.5    &    81.1    &    7.1    &    8.5    &    7    \\ 
  SW0006\_11251.sf    &   18:08:16.2  &   +00:42:06.5   &    28.5961    &    9.9168    &    147.7    &    146.8    &    8.2    &    7.4    &    5    \\ 
  BJ0017\_09541.sf    &   18:10:50.9  &   +22:27:12.4   &    49.2113    &    18.6067    &    82.4    &    70.0    &    7.5    &    5.4    &    4    \\ 
  BJ0017\_093A1.sf    &   18:15:50.8  &   +21:33:42.3   &    48.8172    &    17.1936    &    187.1    &    76.9    &    7.5    &    3.4    &    10    \\ 
  BJ0022\_04071.sf    &   18:18:34.5  &   +19:56:02.7   &    47.5212    &    15.9633    &    177.1    &    84.4    &    7.7    &    9.4    &    7    \\ 
  BJ0018\_052A1.sf    &   18:27:41.1  &   -51:27:44.3   &    343.4328    &    -17.4159    &    217.9    &    89.1    &    7.5    &    11.0    &    10    \\ 
  SW0004\_07481.sf    &   18:30:56.3  &   -38:42:56.3   &    355.7999    &    -12.8951    &    169.6    &    118.1    &    7.2    &    4.2    &    8    \\ 
  BJ0008\_09141.sf    &   18:32:28.8  &   -43:38:08.4   &    351.2853    &    -15.1554    &    196.7    &    101.2    &    7.2    &    11.8    &    4    \\ 
  BJ0020\_052A1.sf    &   18:47:53.4  &   -46:35:41.6   &    349.4568    &    -18.7726    &    105.0    &    81.8    &    7.9    &    3.0    &    10    \\ 
  BJ0011\_00251.sf    &   18:51:02.7  &   -37:19:48.9   &    358.7352    &    -15.9797    &    255.3    &    95.0    &    7.4    &    5.3    &    5    \\ 
  PM0126\_04121.sf    &   18:52:50.0  &   -06:52:08.4   &    26.9697    &    -3.4349    &    503.9    &    366.5    &    7.5    &    14.8    &    2    \\ 
  BJ0026\_01991.sf    &   19:13:18.1  &   -29:27:26.4   &    8.1672    &    -17.3411    &    170.1    &    86.3    &    7.2    &    5.5    &    9    \\ 
  SW0008\_03371.sf    &   19:19:55.4  &   +01:56:38.0   &    37.9275    &    -5.4400    &    279.8    &    217.9    &    7.2    &    8.9    &    7    \\ 
  PM0141\_023B1.sf    &   19:20:38.5  &   +05:20:34.3   &    41.0327    &    -4.0226    &    259.1    &    254.5    &    7.2    &    14.1    &    11    \\ 
  BJ0020\_10761.sf    &   20:09:08.3  &   +02:37:43.1   &    44.4691    &    -15.9480    &    197.2    &    85.3    &    7.4    &    7.2    &    6    \\ 
  BJ0019\_08691.sf    &   20:11:37.0  &   +00:04:49.8   &    42.3034    &    -17.7995    &    270.3    &    76.7    &    7.2    &    9.5    &    9    \\\hline

  \end{tabular}
  \end{center}
  \end{scriptsize}
  \end{table*}



\bibliographystyle{mnras}
\bibliography{mnras_81} 

\bsp	
\label{lastpage}
\end{document}